\documentclass[aps,superscriptaddress,superbib,twocolumn,groupedaddress]{revtex4}

\usepackage[latin1]{inputenc}
\usepackage[T1]{fontenc}

\usepackage{amsmath,amssymb}
\usepackage{graphicx}
\graphicspath{{fig/}}

\usepackage{units}
\usepackage{journames}

\usepackage[final]{neel}

\begin{document}

\renewcommand\topfraction{0.8}
\renewcommand\bottomfraction{0.7}
\renewcommand\floatpagefraction{0.7}

\def\Htc{H_{t\mathrm{,c}}}

\title{Perpendicular anisotropy of ultrathin epitaxial cobalt films on graphene}%

\author{Chi Vo-Van}
\affiliation{Institut N\'{E}EL, CNRS \& Universit\'{e}
Joseph Fourier -- BP166 -- F-38042 Grenoble Cedex 9 -- France}%
\author{Zoukaa Kassir-Bodon}
\affiliation{Institut N\'{E}EL, CNRS \& Universit\'{e}
Joseph Fourier -- BP166 -- F-38042 Grenoble Cedex 9 -- France}%
\author{Hongxin Yang}
\affiliation{SPINTEC (UMR8191 CEA/CNRS/UJF/G-INP), CEA Grenoble, INAC, 38054 Grenoble Cedex 9, France}%
\author{Johann Coraux}
\affiliation{Institut N\'{E}EL, CNRS \& Universit\'{e}
Joseph Fourier -- BP166 -- F-38042 Grenoble Cedex 9 -- France}%
\author{Jan Vogel}
\affiliation{Institut N\'{E}EL, CNRS \& Universit\'{e}
Joseph Fourier -- BP166 -- F-38042 Grenoble Cedex 9 -- France}%
\author{Stefania Pizzini}
\affiliation{Institut N\'{E}EL, CNRS \& Universit\'{e}
Joseph Fourier -- BP166 -- F-38042 Grenoble Cedex 9 -- France}%
\author{Pascale Bayle-Guillemaud}
\affiliation{CEA-Grenoble, INAC/SP2M/LEMMA, 17 rue des Martyrs, Grenoble, France}%
\author{Mairbek Chshiev}
\affiliation{SPINTEC (UMR8191 CEA/CNRS/UJF/G-INP), CEA Grenoble, INAC, 38054 Grenoble Cedex 9, France}%
\author{Laurent Ranno}
\affiliation{Institut N\'{E}EL, CNRS \& Universit\'{e}
Joseph Fourier -- BP166 -- F-38042 Grenoble Cedex 9 -- France}%
\author{Val\'{e}rie Santonacci}
\affiliation{Institut N\'{E}EL, CNRS \& Universit\'{e}
Joseph Fourier -- BP166 -- F-38042 Grenoble Cedex 9 -- France}%
\author{Philippe David}
\affiliation{Institut N\'{E}EL, CNRS \& Universit\'{e}
Joseph Fourier -- BP166 -- F-38042 Grenoble Cedex 9 -- France}%
\author{Violaine Salvador}
\affiliation{CEA-Grenoble, INAC/SP2M/LEMMA, 17 rue des Martyrs, Grenoble, France}%
\author{Olivier Fruchart}
\affiliation{Institut N\'{E}EL, CNRS \& Universit\'{e}
Joseph Fourier -- BP166 -- F-38042 Grenoble Cedex 9 -- France}%

\date{\today}%

\begin{abstract}

Graphene is attractive for spintronics due to its long spin life time and high mobility. So far only thick and polycrystalline slabs 
have been used as ferromagnetic electrodes. We report the growth of flat, epitaxial ultrathin Co films on graphene. These display 
perpendicular magnetic anisotropy in the thickness range $\lengthnm{0.5-1}$, which is confirmed by theory. PMA, epitaxy and 
ultrathin thickness bring new perspectives for graphene-based spintronic devices such as the zero-field control of an arbitrary 
magnetization direction, band matching between electrodes and graphene, and interface effects such as Rashba and electric field 
control of magnetism.

\end{abstract}


\maketitle

\vskip 0.5in

\vskip 0.5in


While graphite exfoliation provides flakes of graphene of lateral size limited to at most $\lengthmicron{100}$\cite{bib-NOV2004}, 
the epitaxial synthesis on SiC\cite{bib-EMT2009} or metals\cite{bib-COR2008,bib-SUT2008} allows for the batch and large-area 
availability and processing\cite{bib-TOM2007} of single and multi-layer graphene. Since 2009 the CVD route is no longer restricted 
to supports made of metallic single crystals, but was extended to thin films such as Ni\cite{bib-REI2009,bib-KIM2009a} or 
Cu\cite{bib-LI2009}, in sheets or deposited on various supporting surfaces. The main motivation so far comes from the demonstrated 
possibility to finally strip off the metal support for the use of the bare graphene sheet in applications, \eg concerning electronic 
transport or photovoltaics.

So far devices have relied on electrodes made with standard clean-room facilities, yielding thick and grainy electrodes, and thus a 
poor control of the microstructure, electronic band matching with graphene, and control over magnetism. This remains far beyond the 
state-of-the art surface-science engineering developed for epitaxial metal-on-metal ferromagnetic systems\finalize{Citation?}. In 
this Letter we report the optimization of the epitaxial growth of Au-capped Co ultrathin films on graphene. These are found to 
display perpendicular magnetic anisotropy~(PMA) in the thickness range $\lengthnm{0.5-1}$. Theory reveals an active role of the 
Co/graphene interface in sustaining PMA. The high uniformity of the layers is confirmed by the exceptionally-low 
coercivity~($\unit[2-10]{mT}$) over the entire PMA range, suitable for the reliable control of magnetization via magnetic or 
electric fields. These features open new perspectives for both free-standing or hererostructure-based graphene devices.

The synthesis was conducted in ultra-high vacuum~(base pressure $\unit[\scientific{3}{-11}]{Torr}$)\cite{bib-FRU2007}. The metallic 
layers were grown by pulsed laser deposition~(PLD)\cite{bib-SHE2004} using a Nd-YAG laser with doubled frequency. A 
computer-controlled mask can be moved in front of the wafer for producing wedge-shaped samples. Details can be found in 
\cite{bib-FRU2007}. CVD is performed with ethylene molecules provided by a dosing tube facing the sample, with a partial pressure of 
$\unit[\scientific{1}{-8}]{Torr}$ measured in the chamber. The samples were grown on Sapphire-C wafers supplied by Roditi 
Ltd.~(miscut angle $\angledeg{0.25}$ or $\angledeg{0.03}$ depending on the batch), which were outgassed twice \insitu under UHV at 
\tempdegC{850} during \unit[45]{min}. \insitu scanning tunneling microscopy~(STM-1 Omicron) and reflection high energy electron 
diffraction~(RHEED, Riber \unit[10]{keV}) were used. High resolution transmission electron microscopy (TEM) was conducted \exsitu 
with a JEOL 4000EX setup with an acceleration voltage of $\unit[400]{kV}$. The cross-sectional specimens were thinned by mechanical 
grinding and ion milling using a PIPS system. Magnetization reversal was probed using the Magneto-Optical Kerr Effect~(MOKE). 
Hysteresis loops were gathered at $\unit[11]{Hz}$ with a laser spot of a few microns and incidence $\angledeg{30}$ away from the 
normal to the plane. A commercial MOKE microscope from Evico-magnetics was also used, with an image in the saturation state 
subtracted from all images. In both cases we are essentially sensitive to perpendicular magnetization~(polar MOKE), and the applied 
field is perpendicular to the plane. The extraordinary Hall effect~(EHE) was measured at room temperature in a four-probe geometry. 
For first principle calculations we used the Vienna ab initio simulation package (VASP)\cite{bib-KRE1996} based on the DFT with the 
generalized gradient approximation \cite{bib-PER1992} and projector augmented wave\cite{bib-BLOe1994}. The Magnetic Anisotropy 
Energy~(MAE) was calculated in two steps. First, the Kohn-Sham equations were solved with no spin-orbit interaction allowing for an 
out-of-plane structural relaxation. Then the spin-orbit coupling was included and the total energy of the system was determined as 
function of the orientation of the magnetic moments. The MAE is computed as the difference between the in-plane and out-of-plane 
total energy values.

\begin{figure}
  \begin{center}
  \includegraphics[width=87mm]{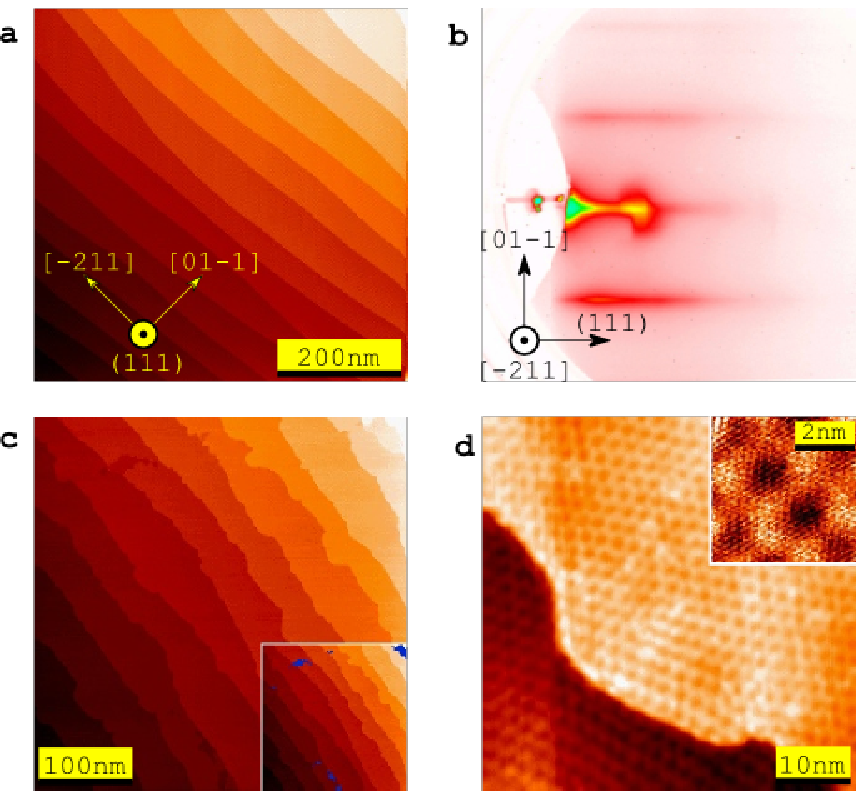}%
  \caption{\label{fig-buffer}(a)\dataref{Fig 1a Ir 600nm CHI11-m6.jpg}~$\unit[600\times600]{nm^2}$
STM topograph of an $\mathrm{Ir}[\lengthnm{8}]$ buffer layer epitaxially grown on Sapphire~C. (b)~RHEED pattern of such an Ir 
surface (c)~\dataref{Fig 1c 400nm CHI16-m18.jpg}~$\unit[400\times400]{nm^2}$ STM topograph of a single sheet of graphene/Ir(111) 
(d)~\dataref{graphene-Ir-5nm Fig4.11d rapport Master Chi.jpg} $\unit[50\times50]{nm^2}$ STM topograph of graphene, with a 
$\unit[5\times5]{nm^2}$ inset with atomic resolution. The superstructure with a $\lengthnm{2.5}$ lattice period is a moir\'{e} pattern, 
resulting from the lattice misfit between the Ir(111) and graphene structures\cite{bib-COR2008}. }
  \end{center}
\end{figure}

The experimental stack is $\mathrm{Au}[\lengthnm{3}](111){/}$ $\mathrm{Co}[t=\lengthnm{0{-}3}](111){/}$ $\mathrm{graphene}{/}$ 
$\mathrm{Ir}[\lengthnm{8}](111){}/$ $\mathrm{Sapphire-C}(0001)$.  Ir is deposited at \tempC{430} and annealed at $\tempC{850}$ 
during \unit[30]{min}. This yields epitaxial Ir(111) with atomically flat terraces whose width is only limited by the miscut angle 
of the wafer\bracketsubfigref{fig-buffer}{a-b}\cite{bib-VVC-tbp}. CVD of graphene on top of Ir is self-limited to a single complete 
layer of graphene similarly to the growth on Ir single crystals\cite{bib-COR2009}\bracketsubfigref{fig-buffer}{c-d}. The graphene 
replicates the atomic smoothness of the Ir(111) buffer layer by extending coherently across its atomic steps.

\begin{figure}
  \begin{center}
  \includegraphics[width=83.515mm]{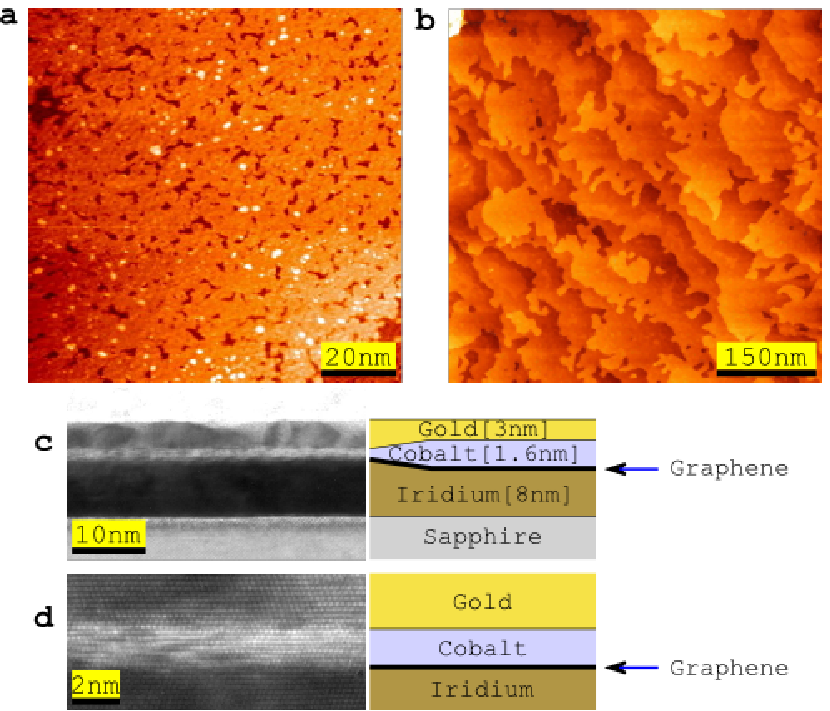}%
  \caption{\label{fig-co}(a)\dataref{4p27mm (4ML).jpg}~$\unit[100\times100]{nm^2}$  STM topograph of $\lengthnm{0.8}$ Co deposited 
at RT on graphene (b)\dataref{Co 600nm annealed 5p5A - FIG5 rapport 2009-12 Chi.jpg}~$\unit[600\times600]{nm^2}$ STM topograph of 
a Co film of thickness $\lengthnm{1.7}$, deposited at RT and annealed at $\tempC{400}$. (c-d)~Cross-sectional high-resolution TEM 
of the stack.}
  \end{center}
\end{figure}

The growth at room temperature~(RT) of Co on graphene proceeds nearly perfectly layer-by-layer up to about 
$\lengthnm{1.5}$\bracketsubfigref{fig-co}{a}\cite{bib-pld-mbe}. For larger thicknesses the roughness progressively increases. In 
both cases the annealing of the deposit at $\tempC{400}$ yields a flat Co film, with a mean terrace width again only limited by the 
miscut angle of the wafer\bracketsubfigref{fig-co}{b}. The stacking is finally terminated with a $\thicknm{3}$-thick RT Au deposit. 
TEM confirms the smoothness and uniformity of the Co layer\bracketsubfigref{fig-co}c. The Co layer is mainly hexagonal compact~(its 
stable RT bulk structure) however with locally stacking faults and/or face-centered-cubic~(stable above $\tempC{425}$ in the bulk, 
however often stabilized in nanostructures) crystals\bracketsubfigref{fig-co}h. The graphene sheet cannot be identified on these 
images due to the graphene-Co spacing being similar to that of Co-Co\cite{bib-GIO2008}. The in-plane epitaxial relationships is 
$\saphir[11{-2}0]{//} \mathrm{Ir}[{-2}11]{// \mathrm{graphene}[1{-1}00]{//} \mathrm{Co}[10{-1}0]}$.


\begin{figure}
  \begin{center}
  \includegraphics[width=86mm]{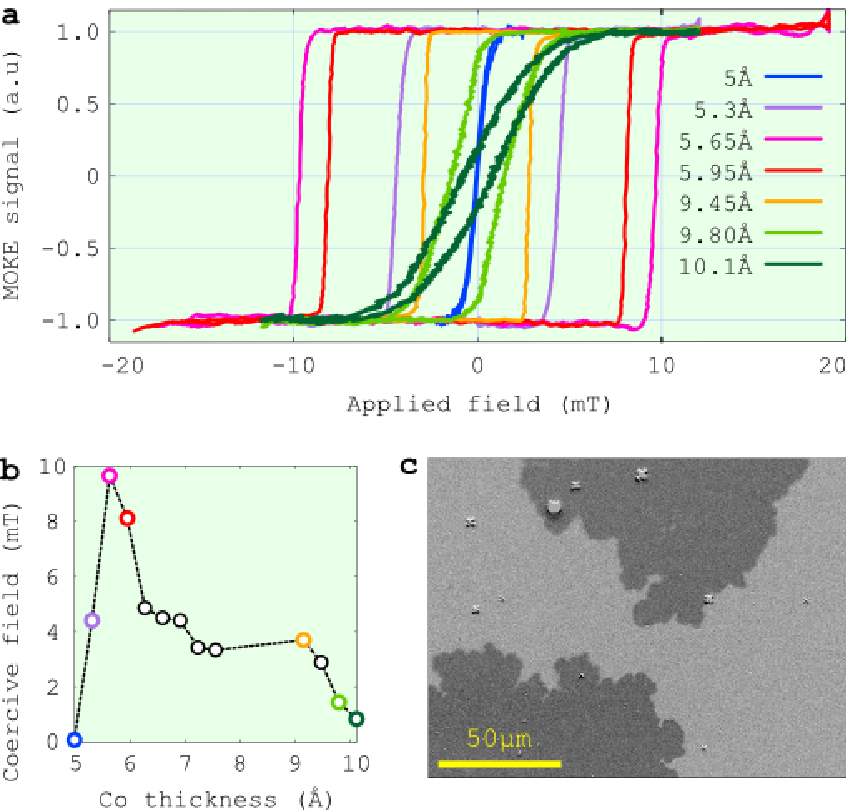}%
  \caption{\label{fig-kerr}(a)\dataref{R\'{e}sum\'{e} CLN2010 Chi}Focused MOKE hysteresis loops of
$\mathrm{Au}[\lengthnm{3}] /\mathrm{Co}[t]/ \mathrm{graphene}/\mathrm{Ir}[\lengthnm{10}]$.  (b)~Coercive field as a function of the 
thickness of the Co layer, as derived from the loops in a. (c)~$\unit[170\times130]{\micron^2}$ MOKE microscopy of
$\mathrm{Au}[\lengthnm{3}] /\mathrm{Co}[\lengthnm{0.8}]/ \mathrm{graphene}/ \mathrm{Ir}$ with a field of \unit[3.5]{mT} applied 
opposite to the initial magnetization direction (the initial domain appears bright. A movie  of a magnetization reversal process is 
provided as supplementary material.}
  \end{center}
\end{figure}

Let us first discuss the MAE of these films. In the uniaxial case the density of MAE is described to first order as 
$E=K\sin^2\theta$ with $K$ in units of $\unit[]{J/m^3}$. With $\theta$ the angle between the magnetization and the normal to the 
film, positive values of $K$ mean perpendicular magnetic anisotropy~(PMA), while negative values mean alignment of magnetization in 
the plane. Our hysteresis loops provide evidence for PMA in the range of thickness 
$\lengthnm{t\approx0.5-1}$\bracketsubfigref{fig-kerr}a. Let us recall that in most magnetic films the magnetization is strongly 
constrained to lie in-the-plane, due to the negative contribution of the magnetostatic energy $E_\mathrm{Shape}=-(1/2)\muZero\Ms^2$ 
to the MAE. PMA may be achieved only for selected cases where the magnetostatic energy is overcome by positive contributions to the 
MAE. In ultrathin films PMA may result from interface and/or magneto-elastic effects\cite{bib-perp-mc}. As both terms decay 
essentially like
$1/t$ with $t$ the film thickness, PMA is restricted to thicknesses typically below
$\lengthnm{1-3}$. Known ultrathin stackings with PMA are either all metal-based\cite{bib-JOH1996} or metal-oxide 
based\cite{bib-ROD2009}\finalize{Aussi r\'{e}f\'{e}rence avec MgO?}. In a first approach Au/Co/graphene may be discussed in the view of 
existing data for the Co-Au interface. This interface favors PMA with a magnitude $\unit[\approx0.5]{mJ/m^2}$\cite{bib-JOH1996}, 
such that full PMA is maintained up to
$t\approx\lengthnm{1.2-1.4}$ for two such interfaces. PMA is maintained up to $t\approx\lengthnm{1}$ in our case, so that we expect 
the contribution of the Co/graphene interface to the PMA to be of similar magnitude.

\begin{figure}
  \begin{center}
  \includegraphics[width=86.786mm]{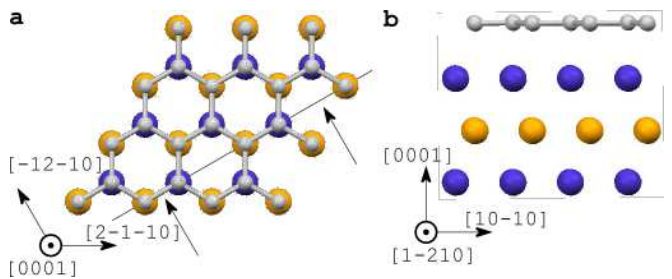}%
  \caption{\label{fig-slabs}Geometry of the Co/graphene slab used for the first principle calculations. (a)~top view and 
(b)~cross-sectional view . The crystallographic indexes refer to the hexagonal notation with three indexes. }
  \end{center}
\end{figure}

We used first-principle calculations to highlight the role in PMA of the Co/graphene, were a significant subtly on strain and local 
environment, the structural details of the Co slab and its interfaces should be taken into account accurately for a quantitative 
discussion. However while Co and graphene have very similar lattice parameters, a large in-plane lattice mismatch exists between Au 
and Co~($\unit[14]{\%}$) and Ir and Co/graphene~($\approx\unit[7]{\%}$). As a consequence a discommensuration moir\'{e} pattern exists 
at both interfaces of the Co/graphene slab, each with a pitch of a few nanometers, and incommensurate one with another. Thus any 
approximant unit cell for the realistic stacking would be at least ten nanometers in lateral size, which is far out of reach of 
first principle calculations. Alternatively, considering highly strained Ir and Au layers to avoid the moir\'{e} would be physically 
unrealistic. With a view to highlighting the physics at the novel Co/graphene interface we compared two simple and tractable cases, 
that of two slabs of thickness three atomic layers of hexagonal compact Co, either free standing or in contact with a commensurate 
graphene sheet on one side\bracketfigref{fig-slabs}. $\thicknm{2}$ vacuum was added on each side with periodic boundary conditions, 
and $21\times21\times1$ $k\;{-}\;\mathrm{points}$ were used. The most stable structural arrangement with graphene was found to be 
that with carbon atoms sitting right atop the Co atoms of the uppermost layer\bracketfigref{fig-slabs}, with an interface distance 
of $\thicknm{0.21}$ after out-of-plane relaxation, consistent with existing results\cite{bib-GIO2008}. Both systems exhibit a strong 
PMA, equalling $\unit[\approx0.845]{mJ/m^2}$ for the free-standing slab, and $\unit[\approx0.828]{mJ/m^2}$ for the graphene-capped 
slab; both figures include the magnetostatic contribution amounting to $\unit[\approx-0.7]{mJ/m^2}$ for this thickness. This means 
that in this ultrathin range the Co/graphene interface plays an active role in promoting PMA, with a strength of the order of 
$\unit[\approx0.75]{mJ/m^2}$, of a similar order of magnitude than that expected from the crude arguments given in the previous 
paragraph.

We now discuss magnetization reversal. The coercive field $\muZero\Hc$ is of the order of a few mT~(\subfigref{fig-kerr}{a-b}). This 
is several orders of magnitude smaller than the anisotropy field, whose lower bound was estimated to $\unit[0.5]{T}$ using EHE. This 
suggests\cite{bib-GIV2003} that magnetization reversal proceeds by the nucleation of a few reversed domains at defects of the 
extended film, followed by an easy propagation of domain walls. The confirmation is gained through the monitoring of magnetic 
domains during magnetization reversal, using Kerr microscopy. Under quasistatic conditions the average size of the domains is larger 
than one hundred micrometers\bracketsubfigref{fig-kerr}c, and magnetization reversal proceeds solely through the propagation of 
domain walls~(see also the Kerr movie as supplementary material, real-time, $\unit[130\times170]{\micron^2}$). Low coercivity is 
usually difficult to achieve in PMA materials because their MAE is large by nature. Similar weak pinning and has been demonstrated 
in selected metal-on-metal systems, however in the very special cases where magnetism is weakened by either selecting extremely low 
thickness\cite{bib-LEE2010} or by weakening anisotropy and magnetism by ion irradiation\cite{bib-CHA1998}. For Co/graphene low 
coercivity is maintained through the entire range of thickness for PMA, which points at the intrinsic quality and homogeneity of the 
layer. \finalize{OPTION: This quality is also attested by the abrupt reorientation transition from perpendicular to in-plane easy 
axis upon a minute change of thickness around $t=\lengthnm{1}$.}

We finally discuss outlooks. There is no reason why growth and magnetic features would differ for other sources of graphene, \eg on 
SiC\cite{bib-EMT2009}, exfoliated\cite{bib-NOV2004} or reported on-metal CVD\cite{bib-REI2009,bib-KIM2009a,bib-LI2009}. Free of a 
conducting buffer layer, these are \apriori better suited for lateral transport devices. For these lateral devices PMA would ensure 
that full remanence, low magnetostatic interactions thanks to the ultrathin thickness, and high thermal stability thanks to the high 
MAE, may be achieved for nanostructures down to very small lateral sizes, required \eg for making use of the short-range RKKY 
coupling\cite{bib-BUN2009a}. Bringing zero-field perpendicular magnetization, PMA also opens the door to the easy realization of 
devices with cross-magnetized electrodes such as those needed for efficient spin-transfer torque magnetization 
precession\finalize{oscillateurs CPP}. Besides, whereas most studied so far considered on spin transport through bare graphene 
between lateral ferromagnetic electrodes, graphene embedded in a metal stack displays specific features whose interest have not been 
exploited yet. For example a graphene/Co/graphene stack embedded within metal weakly bonded with graphene\cite{bib-GIO2008} may 
provide a way to decouple the conduction channels in each layer, and thus ensure a highly spin-polarized conduction channel through 
Co. This may solve the issue of the loss of effective spin polarization in all-metal stacks with PMA investigated for the 
current-induced propagation of domain walls\cite{bib-COR2010}. Other graphene-relevant interfacial effects pertain to novel ways of 
controlling magnetization, such as Rashba spin-orbit coupling\cite{bib-MIR2010} and electric-fields\cite{bib-WEI2007}. The use of 
ultrathin layers is crucial for these interface-based physics so that the relative effects are large. Finally, epitaxial samples are 
also desirable in this case to achieve a good and uniform matching of electronic bands, for engineering the effects and also provide 
simple cases for their fundamental understanding.

To conclude we have developed Au-capped atomically-smooth ultrathin epitaxial Co films on graphene by combining CVD and PLD, which 
display perpendicular magnetic anisotropy~(PMA) in the range of thickness $\lengthnm{0.5-1}$. The availability of such electrodes 
broadens the spectrum of graphene-based devices that can be realized, either for ultra-small lateral sizes or cross-magnetization 
electrodes with PMA, or new interface-based physics in graphene hybrid stackings such as Rashba or electric field control of 
magnetism.

\section*{References}


\begin{thebibliography}{30}
\expandafter\ifx\csname natexlab\endcsname\relax\def\natexlab#1{#1}\fi
\expandafter\ifx\csname bibnamefont\endcsname\relax
  \def\bibnamefont#1{#1}\fi
\expandafter\ifx\csname bibfnamefont\endcsname\relax
  \def\bibfnamefont#1{#1}\fi
\expandafter\ifx\csname citenamefont\endcsname\relax
  \def\citenamefont#1{#1}\fi
\expandafter\ifx\csname url\endcsname\relax
  \def\url#1{\texttt{#1}}\fi
\expandafter\ifx\csname urlprefix\endcsname\relax\def\urlprefix{URL }\fi
\providecommand{\bibinfo}[2]{#2}
\providecommand{\eprint}[2][]{\url{#2}}

\bibitem[{\citenamefont{Novoselov et~al.}(2004)\citenamefont{Novoselov, Geim,
  Morozov, Jiang, Zhang, Dubonos, Grigorieva, and Firsov}}]{bib-NOV2004}
\bibinfo{author}{\bibfnamefont{K.~S.} \bibnamefont{Novoselov}},
  \bibinfo{author}{\bibfnamefont{A.~K.} \bibnamefont{Geim}},
  \bibinfo{author}{\bibfnamefont{S.~V.} \bibnamefont{Morozov}},
  \bibinfo{author}{\bibfnamefont{D.}~\bibnamefont{Jiang}},
  \bibinfo{author}{\bibfnamefont{Y.}~\bibnamefont{Zhang}},
  \bibinfo{author}{\bibfnamefont{S.~V.} \bibnamefont{Dubonos}},
  \bibinfo{author}{\bibfnamefont{I.~V.} \bibnamefont{Grigorieva}},
  \bibnamefont{and} \bibinfo{author}{\bibfnamefont{A.~A.}
  \bibnamefont{Firsov}}, \bibinfo{journal}{\science}
  \textbf{\bibinfo{volume}{306}}, \bibinfo{pages}{666} (\bibinfo{year}{2004}).

\bibitem[{\citenamefont{Emtsev et~al.}(2009)\citenamefont{Emtsev, Bostwick,
  Horn, Jobst, Kellogg, Ley, McChesney, Ohta, Reshanov, Rotenberg, Schmid,
  Waldmann, B.Weber, and Seyller}}]{bib-EMT2009}
\bibinfo{author}{\bibfnamefont{K.~V.} \bibnamefont{Emtsev}},
  \bibinfo{author}{\bibfnamefont{A.}~\bibnamefont{Bostwick}},
  \bibinfo{author}{\bibfnamefont{K.}~\bibnamefont{Horn}},
  \bibinfo{author}{\bibfnamefont{J.}~\bibnamefont{Jobst}},
  \bibinfo{author}{\bibfnamefont{G.~L.} \bibnamefont{Kellogg}},
  \bibinfo{author}{\bibfnamefont{L.}~\bibnamefont{Ley}},
  \bibinfo{author}{\bibfnamefont{J.~L.} \bibnamefont{McChesney}},
  \bibinfo{author}{\bibfnamefont{T.}~\bibnamefont{Ohta}},
  \bibinfo{author}{\bibfnamefont{S.~A.} \bibnamefont{Reshanov}},
  \bibinfo{author}{\bibfnamefont{J.~R.~E.} \bibnamefont{Rotenberg}},
  \bibinfo{author}{\bibfnamefont{A.~K.} \bibnamefont{Schmid}},
  \bibinfo{author}{\bibfnamefont{D.}~\bibnamefont{Waldmann}},
  \bibinfo{author}{\bibfnamefont{H.}~\bibnamefont{B.Weber}}, \bibnamefont{and}
  \bibinfo{author}{\bibfnamefont{T.}~\bibnamefont{Seyller}},
  \bibinfo{journal}{\NatMater} \textbf{\bibinfo{volume}{8}},
  \bibinfo{pages}{203} (\bibinfo{year}{2009}).

\bibitem[{\citenamefont{Coraux et~al.}(2008)\citenamefont{Coraux, {N'Diaye},
  Busse, and Michely}}]{bib-COR2008}
\bibinfo{author}{\bibfnamefont{J.}~\bibnamefont{Coraux}},
  \bibinfo{author}{\bibfnamefont{A.~T.} \bibnamefont{{N'Diaye}}},
  \bibinfo{author}{\bibfnamefont{C.}~\bibnamefont{Busse}}, \bibnamefont{and}
  \bibinfo{author}{\bibfnamefont{T.}~\bibnamefont{Michely}},
  \bibinfo{journal}{\nanolett} \textbf{\bibinfo{volume}{8}},
  \bibinfo{pages}{565} (\bibinfo{year}{2008}).

\bibitem[{\citenamefont{Sutter et~al.}(2008)\citenamefont{Sutter, Flege, and
  Sutter}}]{bib-SUT2008}
\bibinfo{author}{\bibfnamefont{P.~W.} \bibnamefont{Sutter}},
  \bibinfo{author}{\bibfnamefont{J.~I.} \bibnamefont{Flege}}, \bibnamefont{and}
  \bibinfo{author}{\bibfnamefont{E.~A.} \bibnamefont{Sutter}},
  \bibinfo{journal}{\NatMater} \textbf{\bibinfo{volume}{7}},
  \bibinfo{pages}{406} (\bibinfo{year}{2008}).

\bibitem[{\citenamefont{Tombros et~al.}(2007)\citenamefont{Tombros, Jozsa,
  Popinciuc, Jonkman, and {van Wees}}}]{bib-TOM2007}
\bibinfo{author}{\bibfnamefont{N.}~\bibnamefont{Tombros}},
  \bibinfo{author}{\bibfnamefont{C.}~\bibnamefont{Jozsa}},
  \bibinfo{author}{\bibfnamefont{M.}~\bibnamefont{Popinciuc}},
  \bibinfo{author}{\bibfnamefont{H.~T.} \bibnamefont{Jonkman}},
  \bibnamefont{and} \bibinfo{author}{\bibfnamefont{B.~J.} \bibnamefont{{van
  Wees}}}, \bibinfo{journal}{\nature} \textbf{\bibinfo{volume}{448}},
  \bibinfo{pages}{571} (\bibinfo{year}{2007}).

\bibitem[{\citenamefont{Reina et~al.}(2009)\citenamefont{Reina, Jia, Ho,
  Nezich, Son, Bulovic, Dresselhaus, and Kong}}]{bib-REI2009}
\bibinfo{author}{\bibfnamefont{A.}~\bibnamefont{Reina}},
  \bibinfo{author}{\bibfnamefont{X.}~\bibnamefont{Jia}},
  \bibinfo{author}{\bibfnamefont{J.}~\bibnamefont{Ho}},
  \bibinfo{author}{\bibfnamefont{D.}~\bibnamefont{Nezich}},
  \bibinfo{author}{\bibfnamefont{H.}~\bibnamefont{Son}},
  \bibinfo{author}{\bibfnamefont{V.}~\bibnamefont{Bulovic}},
  \bibinfo{author}{\bibfnamefont{M.~S.} \bibnamefont{Dresselhaus}},
  \bibnamefont{and} \bibinfo{author}{\bibfnamefont{J.}~\bibnamefont{Kong}},
  \bibinfo{journal}{\nanolett} \textbf{\bibinfo{volume}{9}},
  \bibinfo{pages}{30} (\bibinfo{year}{2009}).

\bibitem[{\citenamefont{Kim et~al.}(2009)\citenamefont{Kim, Zhao, Jang, Lee,
  Kim, Kim, Ahn, Kim, Choi, and Hong}}]{bib-KIM2009a}
\bibinfo{author}{\bibfnamefont{K.~S.} \bibnamefont{Kim}},
  \bibinfo{author}{\bibfnamefont{Y.}~\bibnamefont{Zhao}},
  \bibinfo{author}{\bibfnamefont{H.}~\bibnamefont{Jang}},
  \bibinfo{author}{\bibfnamefont{S.~Y.} \bibnamefont{Lee}},
  \bibinfo{author}{\bibfnamefont{J.~M.} \bibnamefont{Kim}},
  \bibinfo{author}{\bibfnamefont{K.~S.} \bibnamefont{Kim}},
  \bibinfo{author}{\bibfnamefont{J.-H.} \bibnamefont{Ahn}},
  \bibinfo{author}{\bibfnamefont{P.}~\bibnamefont{Kim}},
  \bibinfo{author}{\bibfnamefont{J.-Y.} \bibnamefont{Choi}}, \bibnamefont{and}
  \bibinfo{author}{\bibfnamefont{B.~H.} \bibnamefont{Hong}},
  \bibinfo{journal}{\nature} \textbf{\bibinfo{volume}{457}},
  \bibinfo{pages}{706} (\bibinfo{year}{2009}).

\bibitem[{\citenamefont{Li et~al.}(2009)\citenamefont{Li, Cai, An, Kim, Nah,
  Yang, Piner, Velamakanni, Jung, Tutuc, Banerjee, Colombo, and
  Ruoff}}]{bib-LI2009}
\bibinfo{author}{\bibfnamefont{X.}~\bibnamefont{Li}},
  \bibinfo{author}{\bibfnamefont{W.}~\bibnamefont{Cai}},
  \bibinfo{author}{\bibfnamefont{J.}~\bibnamefont{An}},
  \bibinfo{author}{\bibfnamefont{S.}~\bibnamefont{Kim}},
  \bibinfo{author}{\bibfnamefont{J.}~\bibnamefont{Nah}},
  \bibinfo{author}{\bibfnamefont{D.}~\bibnamefont{Yang}},
  \bibinfo{author}{\bibfnamefont{R.}~\bibnamefont{Piner}},
  \bibinfo{author}{\bibfnamefont{A.}~\bibnamefont{Velamakanni}},
  \bibinfo{author}{\bibfnamefont{I.}~\bibnamefont{Jung}},
  \bibinfo{author}{\bibfnamefont{E.}~\bibnamefont{Tutuc}},
  \bibinfo{author}{\bibfnamefont{S.~K.} \bibnamefont{Banerjee}},
  \bibinfo{author}{\bibfnamefont{L.}~\bibnamefont{Colombo}}, \bibnamefont{and}
  \bibinfo{author}{\bibfnamefont{R.~S.} \bibnamefont{Ruoff}},
  \bibinfo{journal}{\science} \textbf{\bibinfo{volume}{324}},
  \bibinfo{pages}{1312} (\bibinfo{year}{2009}).

\bibitem[{\citenamefont{Fruchart et~al.}(2007)\citenamefont{Fruchart, Eleoui,
  Jubert, David, Santonacci, Cheynis, Borca, Hasegawa, and
  Meyer}}]{bib-FRU2007}
\bibinfo{author}{\bibfnamefont{O.}~\bibnamefont{Fruchart}},
  \bibinfo{author}{\bibfnamefont{M.}~\bibnamefont{Eleoui}},
  \bibinfo{author}{\bibfnamefont{P.~O.} \bibnamefont{Jubert}},
  \bibinfo{author}{\bibfnamefont{P.}~\bibnamefont{David}},
  \bibinfo{author}{\bibfnamefont{V.}~\bibnamefont{Santonacci}},
  \bibinfo{author}{\bibfnamefont{F.}~\bibnamefont{Cheynis}},
  \bibinfo{author}{\bibfnamefont{B.}~\bibnamefont{Borca}},
  \bibinfo{author}{\bibfnamefont{M.}~\bibnamefont{Hasegawa}}, \bibnamefont{and}
  \bibinfo{author}{\bibfnamefont{C.}~\bibnamefont{Meyer}},
  \bibinfo{journal}{\jpcm} \textbf{\bibinfo{volume}{19}},
  \bibinfo{pages}{053001} (\bibinfo{year}{2007}).

\bibitem[{\citenamefont{Shen et~al.}(2004)\citenamefont{Shen, Gaib, and
  Kirschner}}]{bib-SHE2004}
\bibinfo{author}{\bibfnamefont{J.}~\bibnamefont{Shen}},
  \bibinfo{author}{\bibfnamefont{Z.}~\bibnamefont{Gaib}}, \bibnamefont{and}
  \bibinfo{author}{\bibfnamefont{J.}~\bibnamefont{Kirschner}},
  \bibinfo{journal}{\ssr} \textbf{\bibinfo{volume}{52}}, \bibinfo{pages}{163}
  (\bibinfo{year}{2004}).

\bibitem[{\citenamefont{Kresse and {Furthm\"uller}}(1996)}]{bib-KRE1996}
\bibinfo{author}{\bibfnamefont{G.}~\bibnamefont{Kresse}} \bibnamefont{and}
  \bibinfo{author}{\bibfnamefont{J.}~\bibnamefont{{Furthm\"uller}}},
  \bibinfo{journal}{\prb} \textbf{\bibinfo{volume}{54}}, \bibinfo{pages}{11169}
  (\bibinfo{year}{1996}).

\bibitem[{\citenamefont{Perdew et~al.}(1992)\citenamefont{Perdew, Chevary,
  Vosko, Jackson, Pederson, Singh, and Fiolhais}}]{bib-PER1992}
\bibinfo{author}{\bibfnamefont{J.~P.} \bibnamefont{Perdew}},
  \bibinfo{author}{\bibfnamefont{J.~A.} \bibnamefont{Chevary}},
  \bibinfo{author}{\bibfnamefont{S.~H.} \bibnamefont{Vosko}},
  \bibinfo{author}{\bibfnamefont{K.~A.} \bibnamefont{Jackson}},
  \bibinfo{author}{\bibfnamefont{M.~R.} \bibnamefont{Pederson}},
  \bibinfo{author}{\bibfnamefont{D.~J.} \bibnamefont{Singh}}, \bibnamefont{and}
  \bibinfo{author}{\bibfnamefont{C.}~\bibnamefont{Fiolhais}},
  \bibinfo{journal}{\prb} \textbf{\bibinfo{volume}{46}}, \bibinfo{pages}{6671}
  (\bibinfo{year}{1992}).

\bibitem[{\citenamefont{{Bl\"ochl}}(1994)}]{bib-BLOe1994}
\bibinfo{author}{\bibfnamefont{P.~E.} \bibnamefont{{Bl\"ochl}}},
  \bibinfo{journal}{Phys. Rev. B} \textbf{\bibinfo{volume}{50}},
  \bibinfo{pages}{17953} (\bibinfo{year}{1994}).

\bibitem[{\citenamefont{Vo-Van et~al.}()\citenamefont{Vo-Van, Coraux, and
  Fruchart}}]{bib-VVC-tbp}
\bibinfo{author}{\bibfnamefont{C.}~\bibnamefont{Vo-Van}},
  \bibinfo{author}{\bibfnamefont{J.}~\bibnamefont{Coraux}}, \bibnamefont{and}
  \bibinfo{author}{\bibfnamefont{O.}~\bibnamefont{Fruchart}},
  \bibinfo{note}{unpublished}.

\bibitem[{\citenamefont{Coraux et~al.}(2009)\citenamefont{Coraux, {T N'Diaye},
  Engler, Busse, Wall, Buckanie, {Meyer zu Heringdorf}, {van Gastel}, Poelsema,
  and Michely}}]{bib-COR2009}
\bibinfo{author}{\bibfnamefont{J.}~\bibnamefont{Coraux}},
  \bibinfo{author}{\bibfnamefont{A.}~\bibnamefont{{T N'Diaye}}},
  \bibinfo{author}{\bibfnamefont{M.}~\bibnamefont{Engler}},
  \bibinfo{author}{\bibfnamefont{C.}~\bibnamefont{Busse}},
  \bibinfo{author}{\bibfnamefont{D.}~\bibnamefont{Wall}},
  \bibinfo{author}{\bibfnamefont{N.}~\bibnamefont{Buckanie}},
  \bibinfo{author}{\bibfnamefont{F.-J.} \bibnamefont{{Meyer zu Heringdorf}}},
  \bibinfo{author}{\bibfnamefont{R.}~\bibnamefont{{van Gastel}}},
  \bibinfo{author}{\bibfnamefont{B.}~\bibnamefont{Poelsema}}, \bibnamefont{and}
  \bibinfo{author}{\bibfnamefont{T.}~\bibnamefont{Michely}},
  \bibinfo{journal}{\njp} \textbf{\bibinfo{volume}{11}},
  \bibinfo{pages}{023006} (\bibinfo{year}{2009}).

\bibitem[{bib({\natexlab{a}})}]{bib-pld-mbe}
\bibinfo{note}{At RT the reported growth on graphene/Ir of transition metals
  such as Fe\cite{bib-NDI2009} or Ni\cite{bib-SIC2010} proceeds in the form of
  dilute assemblies of multi-layered clusters. Only deposits thicker than
  several atomic layers or even nanometers monolayers are expected to yield a
  continuous film. Here instead of MBE we use PLD, whose higher instantaneous
  deposition rate and thus higher nucleation rate, is known to favor
  layer-by-layer growth\cite{bib-SHE2004,bib-FRU2010b}.}

\bibitem[{\citenamefont{Giovannetti et~al.}(2008)\citenamefont{Giovannetti,
  Khomyakov, Brocks, Karpan, van~den Brink, and Kelly}}]{bib-GIO2008}
\bibinfo{author}{\bibfnamefont{G.}~\bibnamefont{Giovannetti}},
  \bibinfo{author}{\bibfnamefont{P.~A.} \bibnamefont{Khomyakov}},
  \bibinfo{author}{\bibfnamefont{G.}~\bibnamefont{Brocks}},
  \bibinfo{author}{\bibfnamefont{V.~M.} \bibnamefont{Karpan}},
  \bibinfo{author}{\bibfnamefont{J.}~\bibnamefont{van~den Brink}},
  \bibnamefont{and} \bibinfo{author}{\bibfnamefont{P.~J.} \bibnamefont{Kelly}},
  \bibinfo{journal}{\prl} \textbf{\bibinfo{volume}{101}},
  \bibinfo{pages}{026803} (\bibinfo{year}{2008}).

\bibitem[{bib({\natexlab{b}})}]{bib-perp-mc}
\bibinfo{note}{Here we leave aside those cases where positive contributions to
  the MAE arise from bulk magnetocrystalline anisotropy such as for
  rare-earth-3d, FePt, CoPt}.

\bibitem[{\citenamefont{Johnson et~al.}(1996)\citenamefont{Johnson, Bloemen,
  {den Broeder}, and {de Vries}}}]{bib-JOH1996}
\bibinfo{author}{\bibfnamefont{M.~T.} \bibnamefont{Johnson}},
  \bibinfo{author}{\bibfnamefont{P.~J.~H.} \bibnamefont{Bloemen}},
  \bibinfo{author}{\bibfnamefont{F.~J.~A.} \bibnamefont{{den Broeder}}},
  \bibnamefont{and} \bibinfo{author}{\bibfnamefont{J.~J.} \bibnamefont{{de
  Vries}}}, \bibinfo{journal}{\rpp} \textbf{\bibinfo{volume}{59}},
  \bibinfo{pages}{1409} (\bibinfo{year}{1996}).

\bibitem[{\citenamefont{Rodmacq et~al.}(2009)\citenamefont{Rodmacq, Manchon,
  Ducruet, Auffret, and Dieny}}]{bib-ROD2009}
\bibinfo{author}{\bibfnamefont{B.}~\bibnamefont{Rodmacq}},
  \bibinfo{author}{\bibfnamefont{A.}~\bibnamefont{Manchon}},
  \bibinfo{author}{\bibfnamefont{C.}~\bibnamefont{Ducruet}},
  \bibinfo{author}{\bibfnamefont{S.}~\bibnamefont{Auffret}}, \bibnamefont{and}
  \bibinfo{author}{\bibfnamefont{B.}~\bibnamefont{Dieny}},
  \bibinfo{journal}{\prb} \textbf{\bibinfo{volume}{79}},
  \bibinfo{pages}{024423} (\bibinfo{year}{2009}).

\bibitem[{\citenamefont{Givord et~al.}(2003)\citenamefont{Givord, Rossignol,
  and Barthem}}]{bib-GIV2003}
\bibinfo{author}{\bibfnamefont{D.}~\bibnamefont{Givord}},
  \bibinfo{author}{\bibfnamefont{M.}~\bibnamefont{Rossignol}},
  \bibnamefont{and} \bibinfo{author}{\bibfnamefont{V.~M. T.~S.}
  \bibnamefont{Barthem}}, \bibinfo{journal}{\jmmm}
  \textbf{\bibinfo{volume}{258-259}}, \bibinfo{pages}{1}
  (\bibinfo{year}{2003}).

\bibitem[{\citenamefont{Lee et~al.}(2010)\citenamefont{Lee, Kim, Ryu, Moon,
  Yun, Gim, Lee, Shin, Lee, and Choe}}]{bib-LEE2010}
\bibinfo{author}{\bibfnamefont{J.-C.} \bibnamefont{Lee}},
  \bibinfo{author}{\bibfnamefont{K.-J.} \bibnamefont{Kim}},
  \bibinfo{author}{\bibfnamefont{J.}~\bibnamefont{Ryu}},
  \bibinfo{author}{\bibfnamefont{K.-W.} \bibnamefont{Moon}},
  \bibinfo{author}{\bibfnamefont{S.-J.} \bibnamefont{Yun}},
  \bibinfo{author}{\bibfnamefont{G.-H.} \bibnamefont{Gim}},
  \bibinfo{author}{\bibfnamefont{K.-S.} \bibnamefont{Lee}},
  \bibinfo{author}{\bibfnamefont{K.-H.} \bibnamefont{Shin}},
  \bibinfo{author}{\bibfnamefont{H.-W.} \bibnamefont{Lee}}, \bibnamefont{and}
  \bibinfo{author}{\bibfnamefont{S.-B.} \bibnamefont{Choe}}
  (\bibinfo{year}{2010}).

\bibitem[{\citenamefont{Chappert et~al.}(1998)\citenamefont{Chappert, Barnas,
  Ferr\'{e}, Kottler, Jamet, Chen, Cambril, Devolder, Rousseaux, Mathet, and
  Launois}}]{bib-CHA1998}
\bibinfo{author}{\bibfnamefont{C.}~\bibnamefont{Chappert}},
  \bibinfo{author}{\bibfnamefont{H.}~\bibnamefont{Barnas}},
  \bibinfo{author}{\bibfnamefont{J.}~\bibnamefont{Ferr\'{e}}},
  \bibinfo{author}{\bibfnamefont{V.}~\bibnamefont{Kottler}},
  \bibinfo{author}{\bibfnamefont{J.-P.} \bibnamefont{Jamet}},
  \bibinfo{author}{\bibfnamefont{Y.}~\bibnamefont{Chen}},
  \bibinfo{author}{\bibfnamefont{E.}~\bibnamefont{Cambril}},
  \bibinfo{author}{\bibfnamefont{T.}~\bibnamefont{Devolder}},
  \bibinfo{author}{\bibfnamefont{F.}~\bibnamefont{Rousseaux}},
  \bibinfo{author}{\bibfnamefont{V.}~\bibnamefont{Mathet}}, \bibnamefont{and}
  \bibinfo{author}{\bibfnamefont{H.}~\bibnamefont{Launois}},
  \bibinfo{journal}{\science} \textbf{\bibinfo{volume}{280}},
  \bibinfo{pages}{1919} (\bibinfo{year}{1998}).

\bibitem[{\citenamefont{Bunder and Lin}(2009)}]{bib-BUN2009a}
\bibinfo{author}{\bibfnamefont{J.~E.} \bibnamefont{Bunder}} \bibnamefont{and}
  \bibinfo{author}{\bibfnamefont{H.-H.} \bibnamefont{Lin}},
  \bibinfo{journal}{\prb} \textbf{\bibinfo{volume}{80}},
  \bibinfo{pages}{153414} (\bibinfo{year}{2009}).

\bibitem[{\citenamefont{Cormier et~al.}(2010)\citenamefont{Cormier, Mougin,
  Ferr\'{e}, Thiaville, Charpentier, Pi\'{e}chon, Weil, Baltz, and
  Rodmacq}}]{bib-COR2010}
\bibinfo{author}{\bibfnamefont{M.}~\bibnamefont{Cormier}},
  \bibinfo{author}{\bibfnamefont{A.}~\bibnamefont{Mougin}},
  \bibinfo{author}{\bibfnamefont{J.}~\bibnamefont{Ferr\'{e}}},
  \bibinfo{author}{\bibfnamefont{A.}~\bibnamefont{Thiaville}},
  \bibinfo{author}{\bibfnamefont{N.}~\bibnamefont{Charpentier}},
  \bibinfo{author}{\bibfnamefont{F.}~\bibnamefont{Pi\'{e}chon}},
  \bibinfo{author}{\bibfnamefont{R.}~\bibnamefont{Weil}},
  \bibinfo{author}{\bibfnamefont{V.}~\bibnamefont{Baltz}}, \bibnamefont{and}
  \bibinfo{author}{\bibfnamefont{B.}~\bibnamefont{Rodmacq}},
  \bibinfo{journal}{\prb} \textbf{\bibinfo{volume}{81}},
  \bibinfo{pages}{024407} (\bibinfo{year}{2010}).

\bibitem[{\citenamefont{Miron et~al.}(2010)\citenamefont{Miron, Gaudin,
  Auffret, Rodmacq, Schuhl, Pizzini, Vogel, and Gambardella}}]{bib-MIR2010}
\bibinfo{author}{\bibfnamefont{I.~M.} \bibnamefont{Miron}},
  \bibinfo{author}{\bibfnamefont{G.}~\bibnamefont{Gaudin}},
  \bibinfo{author}{\bibfnamefont{S.}~\bibnamefont{Auffret}},
  \bibinfo{author}{\bibfnamefont{B.}~\bibnamefont{Rodmacq}},
  \bibinfo{author}{\bibfnamefont{A.}~\bibnamefont{Schuhl}},
  \bibinfo{author}{\bibfnamefont{S.}~\bibnamefont{Pizzini}},
  \bibinfo{author}{\bibfnamefont{J.}~\bibnamefont{Vogel}}, \bibnamefont{and}
  \bibinfo{author}{\bibfnamefont{P.}~\bibnamefont{Gambardella}},
  \bibinfo{journal}{\NatMater} \textbf{\bibinfo{volume}{9}},
  \bibinfo{pages}{230} (\bibinfo{year}{2010}).

\bibitem[{\citenamefont{Weisheit et~al.}(2007)\citenamefont{Weisheit, F\"{a}hler,
  Marty, Souche, Poinsignon, and Givord}}]{bib-WEI2007}
\bibinfo{author}{\bibfnamefont{M.}~\bibnamefont{Weisheit}},
  \bibinfo{author}{\bibfnamefont{S.}~\bibnamefont{F\"{a}hler}},
  \bibinfo{author}{\bibfnamefont{A.}~\bibnamefont{Marty}},
  \bibinfo{author}{\bibfnamefont{Y.}~\bibnamefont{Souche}},
  \bibinfo{author}{\bibfnamefont{C.}~\bibnamefont{Poinsignon}},
  \bibnamefont{and} \bibinfo{author}{\bibfnamefont{D.}~\bibnamefont{Givord}},
  \bibinfo{journal}{\science} \textbf{\bibinfo{volume}{315}},
  \bibinfo{pages}{349} (\bibinfo{year}{2007}).

\bibitem[{\citenamefont{N'Diaye et~al.}(2009)\citenamefont{N'Diaye, Gerber,
  Busse, Myslivecek, Coraux, and Michely}}]{bib-NDI2009}
\bibinfo{author}{\bibfnamefont{A.~T.} \bibnamefont{N'Diaye}},
  \bibinfo{author}{\bibfnamefont{T.}~\bibnamefont{Gerber}},
  \bibinfo{author}{\bibfnamefont{C.}~\bibnamefont{Busse}},
  \bibinfo{author}{\bibfnamefont{J.}~\bibnamefont{Myslivecek}},
  \bibinfo{author}{\bibfnamefont{J.}~\bibnamefont{Coraux}}, \bibnamefont{and}
  \bibinfo{author}{\bibfnamefont{T.}~\bibnamefont{Michely}},
  \bibinfo{journal}{\njp} \textbf{\bibinfo{volume}{11}},
  \bibinfo{pages}{103045} (\bibinfo{year}{2009}).

\bibitem[{\citenamefont{Sicot et~al.}(2010)\citenamefont{Sicot, Bouvron,
  Zander, R\"{u}diger, Dedkov, and Fonin}}]{bib-SIC2010}
\bibinfo{author}{\bibfnamefont{M.}~\bibnamefont{Sicot}},
  \bibinfo{author}{\bibfnamefont{S.}~\bibnamefont{Bouvron}},
  \bibinfo{author}{\bibfnamefont{O.}~\bibnamefont{Zander}},
  \bibinfo{author}{\bibfnamefont{U.}~\bibnamefont{R\"{u}diger}},
  \bibinfo{author}{\bibfnamefont{Y.~S.} \bibnamefont{Dedkov}},
  \bibnamefont{and} \bibinfo{author}{\bibfnamefont{M.}~\bibnamefont{Fonin}},
  \bibinfo{journal}{\apl} \textbf{\bibinfo{volume}{96}},
  \bibinfo{pages}{093115} (\bibinfo{year}{2010}).

\bibitem[{\citenamefont{Clavero et~al.}(2010)\citenamefont{Clavero, Cebollada,
  Armelles, and Fruchart}}]{bib-FRU2010b}
\bibinfo{author}{\bibfnamefont{C.}~\bibnamefont{Clavero}},
  \bibinfo{author}{\bibfnamefont{A.}~\bibnamefont{Cebollada}},
  \bibinfo{author}{\bibfnamefont{G.}~\bibnamefont{Armelles}}, \bibnamefont{and}
  \bibinfo{author}{\bibfnamefont{O.}~\bibnamefont{Fruchart}},
  \bibinfo{journal}{\jmmm} \textbf{\bibinfo{volume}{322}}, \bibinfo{pages}{651}
  (\bibinfo{year}{2010}).

\end{thebibliography}

\section*{Acknowledgements}

We thank J. Debray for wafer orientation. C.V.V, H.Y and M.S. acknowledge financial support from Fondation Nanosciences. This work 
was partially supported by ANR-07-NANO-034 \textsl{Dynawall}.

\end{document}